# Diversity-Aware Batch-Mode Active Learning for Efficient Sampling in Data-Driven Constitutive Modeling

Ronak Shoghi[1*], Lukas Morand[2], Dirk Helm[2], Alexander Hartmaier[1]

[1] ICAMS, Ruhr University Bochum, Germany

[2] Fraunhofer Institute for Mechanics of Materials IWM, Freiburg, Germany

## Abstract

The constitutive behavior of materials is modeled through relationships between stress, strain, and possibly additional internal variables. This results in relatively high-dimensional feature spaces for machine learning models rendering the efficient generation of informative datasets essential as brute force methods suffer from the curse of dimensionality. This work introduces a diversity-aware batch-mode query-by-committee active-learning strategy to generate datasets of maximum information content at minimum cost. In contrast to existing methods, this novel method selects multiple informative, non-redundant queries per iteration, enabling concurrent generation of informative datasets and reducing the number of machine-learning retraining cycles. A central component of this method is a cosine-similarity-based metric that complements the uncertainty criterion based on committee variance by promoting within-batch diversity. The query selection is guided by committee variance and a diversity-promoting criterion. The approach is benchmarked for efficient stress-space sampling in data-driven constitutive modeling. In this setting, a committee of support vector classifiers approximates the so-called yield surface, which is a manifold dividing the six-dimensional stress space into an elastic and plastic domain. We demonstrate that the method handles different batch sizes robustly, maintains high within-batch diversity, and rapidly reduces committee uncertainty. The resulting machine-learning yield surfaces achieve predictive accuracy comparable to sequential active learning, while requiring substantially fewer retraining cycles. This makes the proposed approach an efficient strategy for stress-space sampling in data-driven constitutive modeling and for reducing time-to-solution via concurrent data collection in each iteration.

## 1. Introduction

In solid mechanics, constitutive models describe material response under mechanical loading through relationships between stress, strain and, where relevant, additional internal variables. The central element of constitutive models for elastoplastic material behavior is the yield function. This yield function subdivides the six-dimensional stress space into an elastic domain, where it takes negative values and a plastic domain where it is equal to (cf. classical plasticity models) or larger than zero (cf. Perzyna-type [1] viscoplasticity models). Classical phenomenological formulations describe this yield function through analytical expressions based on an equivalent stress or, more generally, linear transformations of the stress tensor: Examples for such yield functions include the isotropic yield functions introduced by von Mises [2] and Tresca [3] and the anisotropic yield functions by Hill [4] and Barlat [5]–[7]. Despite their differences, these models share a common limitation: their functional structure is prescribed a priori and calibrated through material-specific parameters, whose identification can be demanding and non-unique, while the resulting formulations often become increasingly complex for anisotropic behavior. These limitations, have motivated emergence of data-driven approaches for constitutive modeling to learn material behavior directly from data, see [8] for an overview. Early contributions of Ortiz [9] and Eggersman [10] in a model-free approach eliminate explicit constitutive equations by solving boundary-value problems directly on material datasets. In contrast, ML-based approaches learn an explicit functional representation of the constitutive response from data, ranging from black-box neural networks [11] and recurrent architectures for history-dependent behavior [12], [13], to structure-aware models enforcing tensorial invariance [14] and physics-/thermodynamics-informed formulations[15]–[18]. Regardless of the specific framework, the predictive performance of a data-driven constitutive model depends on the availability of representative training data in the relevant constitutive state space.

Data-driven alternatives to classical yield functions have been widely investigated, with ground-truth data commonly generated using different sampling strategies in multidimensional stress space or its invariant-based representations. Existing studies have, for example, used discretization of invariant stress space through prescribed Lode angles or triaxiality values [19], [20], uniform sampling of the six-dimensional stress space [21], random sampling in stress space [22], and prescribed loading directions in reduced biaxial principal stress space stress spaces [23]. Despite differences in parameterization and modeling framework, these static approaches rely on predefined sampling strategies in stress space or its invariant-based representations. In high-dimensional stress spaces, such static sampling approaches can lead to unnecessarily many evaluations in well-resolved regions and insufficient resolution in more complex areas of the stress space. Consequently, the efficiency of ground-truth data generation becomes a critical aspect, particularly when yield stresses are obtained from computationally expensive virtual experiments.

To address the inefficiency of static sampling approaches, active learning treats ground-truth data generation as an adaptive and model-guided process rather than a static preprocessing step. In contrast to conventional supervised learning, where the training dataset is fixed a priori, active learning integrates sampling into the training loop. An initial surrogate model is first trained and then used to identify new locations in the feature space where additional ground-truth evaluations are expected to be most informative. These newly queried data points are added to the training set, and the model is retrained iteratively. Thus, data acquisition and model construction are coupled in a closed-loop process that seeks to maximize information gain while minimizing expensive evaluations. Active learning methods are distinguished primarily by their query strategy, i.e., how the informativeness of candidate inputs is quantified. See [24]–[26] for an overview. Uncertainty sampling [27] queries points where the current model is least confident. Expected model change or expected error reduction strategies [28] choose points predicted to most update the model or most reduce future generalization error. Query-by-Committee (QBC) [29] defines the acquisition criterion directly in terms of disagreement in feature space. Regions where independently trained models yield divergent predictions correspond to areas of insufficient data coverage. Sampling in these regions refines the surrogate locally and reduces uncertainty. Morand et al. [30] demonstrated the effectiveness of the QBC strategy in the materials science domain, particularly by applying it to neural network (NN) models to efficiently explore microstructure and property spaces of materials. Wessel et al. [31] utilized QBC to efficiently sample virtual experiments across the full stress state, which was then applied to determine parameters for anisotropic yield models. Shoghi et al. [32] applied an active learning approach using the QBC algorithm for support vector classification to train an ML yield function [33]. It was shown that the proposed active learning strategy resulted in more efficient sampling of the six-dimensional stress space.

However, active learning strategies are often formulated in a sequential manner, where a single data point is selected, and the surrogate model is retrained after each query. While effective, such an approach can incur substantial retraining overhead, particularly when surrogate training is computationally non-negligible. To address this issue, batch-mode active learning aims to select multiple informative data points per iteration, thereby reducing the number of retraining cycles and improving overall computational efficiency [34]. Typically, a single instance selection criterion cannot be easily extended to select multiple instances. Moreover, selecting a batch requires a detailed investigation into the trade-offs between retraining costs and performance gains [34], [35]. While batch-mode active learning has been studied in broader machine learning contexts, applications to data-driven constitutive modeling remain scarce. To the authors' knowledge, Qu et al. [36] is the only reference describing batch-mode active learning in the constitutive modelling domain. Therein, a committee-based active learning strategy is employed, in which candidate strain paths are ranked according to prediction variance across multiple neural networks, and those with the highest disagreement are selected to generate new simulation data. Although several samples may be added per round, the selection is based purely on variance ranking without enforcing diversity among the chosen points. As noted in their study, the selected samples may cluster in specific regions of the strain space. Such clustering can lead to redundant information within a batch and limit sampling efficiency in high-dimensional settings. This contradicts the actual purpose of implementing active learning strategies.

Motivated by this limitation, in the present work, we propose a diversity-aware batch-mode QBC strategy for efficient training-data generation through adaptive sampling. Although the proposed approach is not restricted to a particular constitutive representation, it is benchmarked here for training machine learning yield functions. Building upon the approach introduced by Shoghi et al. [32], the yield surface is approximated using a support vector classifier (SVC). Within the QBC framework, uncertainty is quantified from the prediction variance across a committee of SVC models, which is used to identify informative queries in uncertain regions of the stress space. The proposed extension augments this variance-based acquisition with a diversity criterion, thereby enabling efficient batch sampling that reduces the number of costly retraining cycles and, when ground-truth evaluations can be parallelizable, allows multiple labels to be generated concurrently. In the benchmark setting considered here, the candidate queries are unit loading directions in stress space, while the corresponding ground-truth data are generated using Hill's anisotropic yield criterion as a reference material model. This enables a systematic assessment of sampling efficiency, measured in terms of retraining effort and time-to-solution, together with within-batch diversity in high-dimensional stress space.

## 2. Methods

This section presents the methodology employed for learning a data-driven yield function using support vector classification (SVC) combined with batch-mode active learning in stress space. First, the formulation of the SVC-based yield function is introduced, where the decision function defines the learned yield surface in stress space. Next, the stress-space representation is described, including the radial parameterization of stress states and the generation of ground-truth data using an anisotropic Hill-type reference material. Finally, the proposed diversity-aware batch-mode active learning strategy is presented, together with the overall workflow and evaluation protocol used for assessing sampling efficiency and model performance.

### 2.1. ML yield function

A yield function $f(\boldsymbol{\sigma})$ classifies material response at a given stress state $\boldsymbol{\sigma}$ into elastic ($f(\boldsymbol{\sigma}) < 0$) or plastic ($f(\boldsymbol{\sigma}) \geq 0$). The zero level-set $f(\boldsymbol{\sigma}) = 0$ defines the yield surface, a geometric object in stress space that separates the two regimes [37], [38]. From a data-driven perspective, yield-surface identification can therefore be cast as a binary classification problem in stress space. Given stress states labeled as elastic or plastic, a support vector classifier (SVC) can be trained such that its decision boundary approximates the yield surface [33].

Support vector classification (SVC) is a supervised learning method based on maximum-margin separation, extended via kernel mappings and soft-margin regularization [39]–[41]. SVC determines a decision boundary that maximizes the margin between two classes while permitting limited violations through a soft-margin formulation. When the labels represent elastic and plastic material states, this boundary can be interpreted as a data-driven yield surface that separates elastic from plastic response. The resulting decision function can be written as:

$$f_{\mathrm{ML}}(\boldsymbol{\sigma}) = \sum_{k=1}^{N_{\mathrm{SV}}} \alpha_k \, y_k K(\boldsymbol{\sigma}_k, \boldsymbol{\sigma}) + b, \qquad K(\boldsymbol{\sigma}_k, \boldsymbol{\sigma}) = \exp(-\gamma \|\boldsymbol{\sigma} - \boldsymbol{\sigma}_k\|^2). \tag{1}$$

Here $\boldsymbol{\sigma}_k$ denote the support vectors, $y_k \in \{-1, +1\}$ are the associated class labels, $\alpha_k$ are the learned coefficients, and $b$ is the bias term. If $N_\mathrm{t}$ is the total number of training data points, only $N_\mathrm{sv} < N_\mathrm{t}$ with $\alpha_k > 0$ act as the support vectors and contribute to the sum. The coefficients $\alpha_k$ and bias term $b$ are obtained by solving the standard SVC dual optimization problem, which promotes a maximum-margin separation between elastic and plastic classes while permitting limited classification errors controlled by the soft-margin parameter $C$. The radial basis function (RBF) $K(\boldsymbol{\sigma}_k, \boldsymbol{\sigma})$ embeds stress states into a high-dimensional Hilbert space [42], enabling nonlinear and flexible separating surfaces [43]. The hyperparameter $\gamma$ determines how far each training point's influence reaches. Smaller values produce a broader, smoother decision boundary, whereas larger values make the model focus more tightly on nearby training points [44], [45]. Training $f_\mathrm{ML}$ requires labeled training data that resolve the elastic–plastic boundary in stress space. Each training sample consists of a stress state $\boldsymbol{\sigma} \in \mathbb{R}^6$ together with a binary label $y \in \{-1, +1\}$ indicating elastic or plastic response. Because the yield surface is a hypersurface in this high-dimensional space, the dataset must provide sufficient coverage of the boundary. In practice, this is achieved by sampling multiple loading directions and generating labeled points in the vicinity of the yield onset along each direction. The following Section 2.2 explains how stress states are defined and sampled, how yield onset is determined using the reference material model, and how the labeled dataset is constructed.

## 2.2. Stress-space representation and ground-truth data generation

To describe yielding under general multiaxial loading, we consider the yield surface as a hypersurface in stress space. Since the Cauchy stress tensor is symmetric and has six independent components, we therefore represent stress states in $\mathbb{R}^6$ using Voigt notation:

$$\boldsymbol{\sigma} = [\sigma_{11}, \sigma_{22}, \sigma_{33}, \sigma_{12}, \sigma_{23}, \sigma_{13}] \in \mathbb{R}^6. \tag{2}$$

Any nonzero stress state can be decomposed into a magnitude and a direction

$$\boldsymbol{\sigma} = \mu\hat{\boldsymbol{\sigma}}, \qquad \mu > 0, \qquad \hat{\boldsymbol{\sigma}} \in S^5 \subset \mathbb{R}^6. \tag{3}$$

where $\hat{\boldsymbol{\sigma}}$ is a unit loading direction, satisfying $\|\hat{\boldsymbol{\sigma}}\| = 1$,and therefore fixes the relative proportions of the stress components and lies on the unit hypersphere $S^5$, while $\mu$ scales the stress magnitude along that direction. For a given $\hat{\boldsymbol{\sigma}}$, yield onset is characterized by the critical amplitude $\mu_\mathrm{y}(\hat{\boldsymbol{\sigma}})$, provided by a ground-truth oracle, meaning a reference yield criterion that returns the yield-onset point for a queried loading direction. In this work, Hill's anisotropic yield criterion is used as this oracle:

$$F(\sigma_{22} - \sigma_{33})^2 + G(\sigma_{33} - \sigma_{11})^2 + H(\sigma_{11} - \sigma_{22})^2 + 2L{\sigma_{23}}^2 + 2M{\sigma_{13}}^2 + 2N{\sigma_{12}}^2 = {\sigma_\mathrm{y}}^2. \tag{4}$$

Here $F, G, H, L, M, N$ are anisotropy parameters and $\sigma_\mathrm{y}$ is the uniaxial yield strength. Their values, listed in Table 1, are adopted from [32]. For each direction $\hat{\boldsymbol{\sigma}}$, $\mu_\mathrm{y}(\hat{\boldsymbol{\sigma}})$ is obtained by scaling $\mu$ in $\boldsymbol{\sigma} = \mu\hat{\boldsymbol{\sigma}}$ until Eq. (4) is satisfied using a one-dimensional root-finding procedure.

Given a set of unit loading directions $\{\hat{\boldsymbol{\sigma}}_n\} \subset S^5$, the corresponding yield-onset points are

$$\boldsymbol{\sigma}_{y,n} = \mu_y(\hat{\boldsymbol{\sigma}}_n)\hat{\boldsymbol{\sigma}}_n. \tag{5}$$

These points form a discrete representation of the yield surface in $\mathbb{R}^6$ and serve as the basis for constructing labeled training data. In this case, for each yield onset point $\boldsymbol{\sigma}_{y,n}$, elastic and plastic stress states are generated by radial scaling along the same loading direction. Stress levels inside the yield surface (e.g., 10–90% of the yield-onset magnitude) are labeled elastic ($y = -1$), whereas stress levels outside the yield surface (e.g., 110–200%) are labeled plastic ($y = +1$). This labeling procedure is illustrated in **Figure 1**. The remaining task is to select the directions $\{\hat{\boldsymbol{\sigma}}_n\}$ on $S^5$ efficiently to obtain an accurate yield-surface representation, which is addressed by the proposed batch-mode active learning strategy in **Section 2.3**.

Table 1 – Material parameters for the analytical Hill anisotropic yield criterion, adopted from Ref.[32].

| Material Parameters | $F$ | $G$ | $H$ | $L$ | $M$ | $N$ | $\sigma_y$(MPa) |
|---|---|---|---|---|---|---|---|
| Parameter Values | 1 | 0.7 | 1.4 | 3.9 | 2.4 | 3 | 50 |

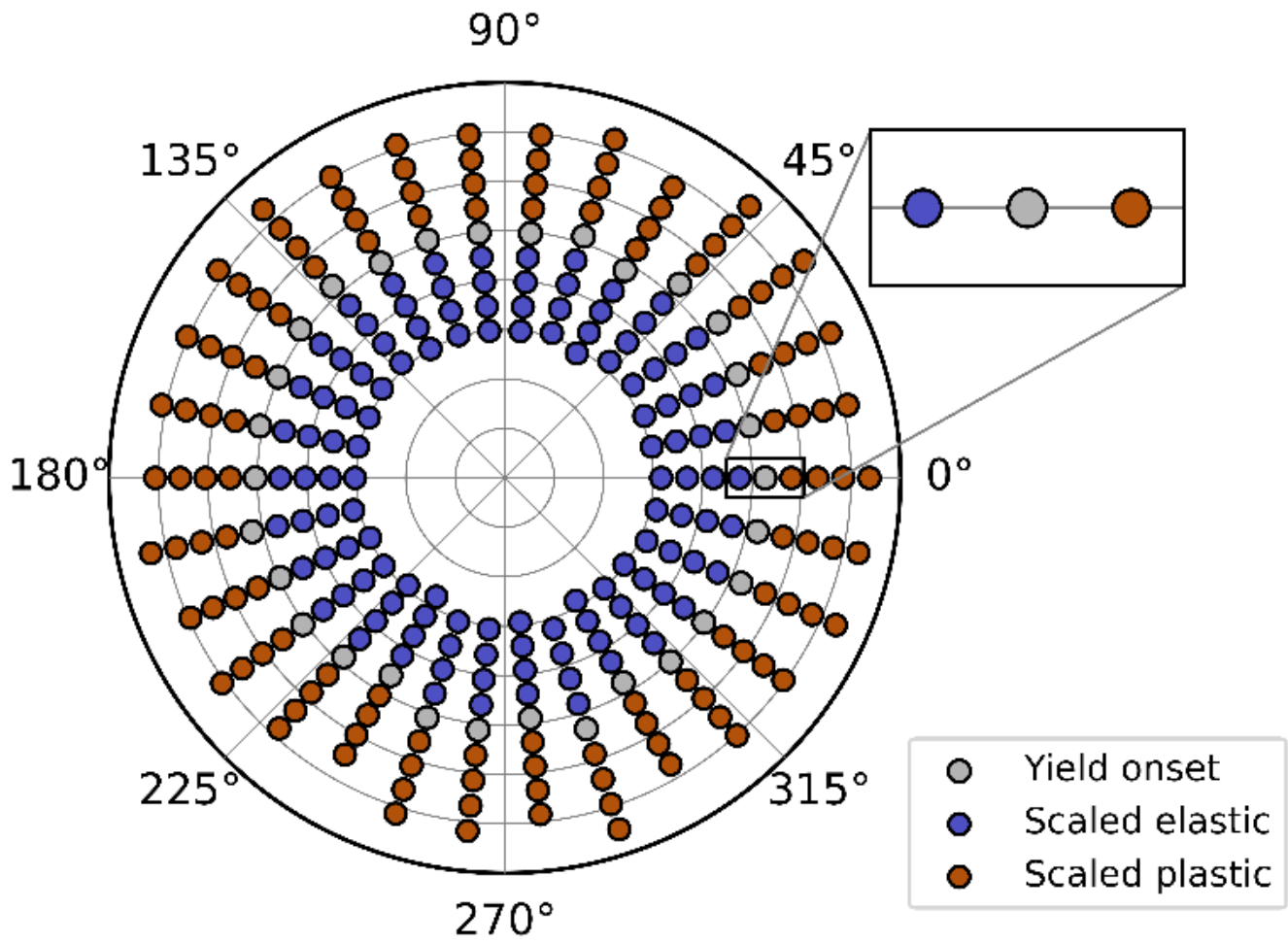

**Figure 1** – Construction of labeled training samples from directional yield-onset points. For each loading direction, the oracle provides a yield-onset stress $\boldsymbol{\sigma}_y$ (gray markers). Labeled training points are then generated by radially scaling $\boldsymbol{\sigma}_y$ along the same direction while keeping the loading direction fixed: stress levels below yield are labeled elastic (blue), and stress levels above yield are labeled plastic (orange). For visualization, yield-onset and scaled points are projected onto a polar plot using loading angle (in the $\pi$-plane) and equivalent stress magnitude. In the dataset used here, four elastic and four plastic scaled samples are generated per direction.

## 2.3. Batch-mode active learning strategy

The anisotropic Hill yield criterion defined as the reference material model serves as the oracle in the present active-learning framework. In the following, we describe how the proposed approach selects the unit loading directions $\{\hat{\boldsymbol{\sigma}}_n\} \subset S^5$ efficiently in a batch-wise manner. We cast direction selection as a batch-mode active learning problem: starting from a small initial set of labeled samples, a committee of SVC models is trained iteratively as explained in **Section 2.1.** Then, a batch of new directions is selected that are expected to be most informative for improving the yield-surface approximation. To quantify informativeness, we use committee-based uncertainty, and to avoid selecting redundant directions within the same iteration, we incorporate an explicit diversity mechanism during batch construction.

We start with a small initial randomly sampled unit loading direction $\mathcal{D} = \{\hat{\boldsymbol{\sigma}}_n\}_{n=1}^{N_0} \subset S^5$. As explained in **Section 2.2**, the oracle is queried for these directions to obtain the corresponding yield-onset points, and labeled elastic/plastic samples are generated by radial scaling. This initial labeled dataset forms the starting point for the active-learning loop. At each iteration, we train a committee of $N$ SVC models on the current labeled dataset. Committee diversity is introduced by training each SVC on a different random 80% split of the current labeled dataset, so that individual committee members see slightly different training data, which induces variation in their learned decision boundaries and enables uncertainty estimation through disagreement [24], [46]. To quantify informativeness for direction selection, we adopt a QBC uncertainty criterion. Let $\hat{\boldsymbol{\sigma}} \in S^5$ denote a candidate unit loading direction. To evaluate uncertainty without querying the oracle, we evaluate the committee predictions at a fixed stress level along this direction, $\boldsymbol{\sigma}_{\text{probe}} = \mu_0 \hat{\boldsymbol{\sigma}}$, where $\mu_0 = 0.5\sigma_{\text{y}}$ is kept constant for all candidates. Each committee member $\eta = 1, \ldots, N$ provides a prediction $f_{\text{ML}}(\boldsymbol{\sigma}^*)$, and committee disagreement is quantified by the prediction variance:

$$\text{Var}(\hat{\boldsymbol{\sigma}}) = \frac{1}{N} \sum_{\eta=1}^{N} \left( f_{\text{ML},\eta}(\boldsymbol{\sigma}_{\text{probe}}) - \bar{f}(\boldsymbol{\sigma}_{\text{probe}}) \right)^2, \tag{6}$$

where $N$ denotes the number of committee members, $f_{\text{ML},\eta}(\boldsymbol{\sigma}_{\text{probe}})$ is the prediction of the $\eta$-th SVC model and $\bar{f}(\boldsymbol{\sigma}_{\text{probe}})$ is the mean prediction over all committee members. Here, $\text{Var}(\hat{\boldsymbol{\sigma}})$ denotes the variance of the committee predictions and therefore provides a scalar measure of committee disagreement for the candidate loading direction $\hat{\boldsymbol{\sigma}}$. A high variance therefore corresponds to high disagreement within the committee for a given $\boldsymbol{\sigma}_{\text{probe}}$. The first direction in each batch is selected by maximizing this disagreement:

$$\hat{\boldsymbol{\sigma}}_1^* = \arg\max_{\hat{\boldsymbol{\sigma}} \in S^5} \text{Var}(\hat{\boldsymbol{\sigma}}). \tag{7}$$

Here, $\arg\max$ returns the loading direction in $S^5$ at which the variance is maximal. In our work, this maximization is solved using the global differential evolution (DE) [47] optimization algorithm. Since DE is a minimization algorithm, we equivalently minimize the negative variance $-\text{Var}(\hat{\boldsymbol{\sigma}})$.

However, if a batch of points were chosen within an active learning cycle by maximizing variance alone, the optimizer returns multiple directions from the same highly uncertain neighborhood of $S^5$, resulting in redundancy. To avoid this, we augment the uncertainty criterion with an additional diversity term that favors angular separation from directions already selected in the current batch. Thus, after choosing the first batch direction $\hat{\boldsymbol{\sigma}}_1^*$ solely based on committee disagreement, the remaining batch elements are selected using a criterion that balances both uncertainty and diversity.

Because loading directions are unit vectors on $S^5$, angular similarity between two directions can be measured by cosine similarity:

$$\cos(\hat{\boldsymbol{\sigma}}, \hat{\boldsymbol{\sigma}}_j^*) = \hat{\boldsymbol{\sigma}} \cdot \hat{\boldsymbol{\sigma}}_j^* \in [-1,1]. \tag{8}$$

Values close to 1 indicate nearly identical directions, whereas smaller values indicate larger angular separation. Cosine-based angular measures have been used previously to describe loading-path changes in constitutive modeling [48], [49]; in the present work, cosine similarity is used instead as a diversity measure for selecting distinct loading directions within each active-learning batch. For the $i$-th element of a batch ($i \geq 2$), the similarity of a candidate direction $\hat{\boldsymbol{\sigma}}$ to the previously selected batch directions $\{\hat{\boldsymbol{\sigma}}_j^*\}_{j=1}^{i-1}$ is quantified by

$$D_i(\hat{\boldsymbol{\sigma}}) = -1 + \sum_{j=1}^{i-1} \left(\hat{\boldsymbol{\sigma}} \cdot \hat{\boldsymbol{\sigma}}_j^*\right). \tag{9}$$

This quantity becomes larger when the candidate direction is strongly aligned with directions already present in the batch and therefore acts as a redundancy measure. The shift by $-1$ ensures that the resulting objective favors candidate directions with high uncertainty and low similarity to the already selected batch directions. Using the diversity term $D_i(\hat{\boldsymbol{\sigma}})$, the remaining batch directions are then selected by minimizing

$$\hat{\boldsymbol{\sigma}}_i^* = \arg\min_{\hat{\boldsymbol{\sigma}} \in S^5} \left(\mathrm{Var}(\hat{\boldsymbol{\sigma}}) D_i(\hat{\boldsymbol{\sigma}})\right), \qquad i = 2, \ldots, b. \tag{10}$$

Here, $\arg\min$ returns the loading direction in $S^5$ at which the combined objective is minimal. For $i = 2$, Eq. (9) becomes

$$D_2(\hat{\boldsymbol{\sigma}}) = -1 + \hat{\boldsymbol{\sigma}} \cdot \hat{\boldsymbol{\sigma}}_1^*, \tag{11}$$

and therefore

$$\mathrm{Var}(\hat{\boldsymbol{\sigma}}) D_2(\hat{\boldsymbol{\sigma}}) = -\,\mathrm{Var}(\hat{\boldsymbol{\sigma}})(1 - \hat{\boldsymbol{\sigma}} \cdot \hat{\boldsymbol{\sigma}}_1^*). \tag{12}$$

Hence, for the second batch element, minimizing Eq. (10) is equivalent to maximizing uncertainty weighted by the angular separation from the first selected batch direction. For $i > 2$, the same mechanism extends through the sum in Eq. (9). After constructing the full batch $\mathcal{B} = \{\hat{\boldsymbol{\sigma}}_1^*, \ldots, \hat{\boldsymbol{\sigma}}_b^*\}$, the oracle is queried for these $b$ directions and the resulting labeled samples are appended to the initial direction set via $\mathcal{D} \leftarrow \mathcal{D} \cup \mathcal{B}$.

## 3. Results and Discussion

The proposed batch-mode active learning strategy is evaluated for batch sizes of 2, 3, and 4, starting from an initial set of $N_0 = 100$ randomly generated unit loading directions $\mathcal{D}^{(0)} = \{\hat{\boldsymbol{\sigma}}_n\}_{n=1}^{100} \subset S^5$. These initial directions are queried with the oracle to obtain yield-onset points and to construct the initial labeled training data, which is used to train the initial committee of $N = 5$ SVC models. Active learning then proceeds under a fixed oracle-query budget of 200 new directions beyond the initialization. Accordingly, the number of iterations is $t = 100$ for batch size 2, $t = 67$ for batch size 3 (201 queries), and $t = 50$ for batch size 4, yielding approximately the same total number of acquired directions across batch sizes. For the SVC model, the Scikit-learn [50] implementation is used, with hyperparameters optimized via 5-fold cross-validation grid search [51]. After each iteration, model performance is evaluated on an independent test set of 300 randomly generated unit loading directions $\{\hat{\boldsymbol{\sigma}}_m\}_{m=1}^{300} \subset S^5$, which are never used during training. For each test direction, reference labels are obtained by querying the oracle and generating elastic/plastic samples via the same radial scaling procedure described in Section 2.2. Performance improvement over active learning iterations is quantified using Matthew's correlation coefficient (MCC). To compute MCC, we use the confusion matrix counts: True Elastic (TE), True Plastic (TP), False Plastic (FP), and False Elastic (FE). For each committee member,

$$\mathrm{MCC} = \frac{(\mathrm{TP})(\mathrm{TE}) - (\mathrm{FP})(\mathrm{FE})}{\sqrt{(\mathrm{TP}+\mathrm{FP})(\mathrm{TP}+\mathrm{FE})(\mathrm{TE}+\mathrm{FP})(\mathrm{TE}+\mathrm{FE})}} \tag{13}$$

We report the committee-average MCC on the unseen test set throughout the active-learning process. All experiments are implemented using pyLabFEA [52] library. To assess the effectiveness of the proposed batch-mode sampling strategy with respect to its two design objectives, we report complementary metrics that track within-batch diversity and uncertainty reduction over the active-learning process. Intra-batch diversity is quantified by the cosine distance between directions selected within the same iteration, while informativeness is measured by the committee prediction variance (disagreement) at the selected queries.

## 3.1. Batch-Mode Sampling for Batch Size 2

**Figure 2 (a)** reports the within-batch diversity achieved by the proposed batch acquisition strategy for batch size $b = 2$ over 100 iterations. In each iteration $t$, the first direction in the batch, $\hat{\boldsymbol{\sigma}}_1{}^{(t)}$ is selected by maximizing committee disagreement (Eq. (7)), while the second direction $\hat{\boldsymbol{\sigma}}_2{}^{(t)}$ is selected by maximizing a joint criterion (Eq. (10)) that promotes both high predictive variance and directional diversity relative to $\hat{\boldsymbol{\sigma}}_1{}^{(t)}$. To quantify intra-batch diversity, the cosine distance is computed as $d_{12}^{(t)} = 1 - \hat{\boldsymbol{\sigma}}_1{}^{(t)} \cdot \hat{\boldsymbol{\sigma}}_2{}^{(t)} \in [0,2]$, where $d = 0$ corresponds to identical load paths, $d = 1$ to orthogonal and $d = 2$ to opposite load paths. The plotted values were averaged over three independent runs (blue markers), and the shaded blue envelope denotes the corresponding ±1 standard-deviation interval.

The mean cosine distance remains substantially above the random-pair baseline ($\approx 1$) throughout the learning process, with an overall mean of 1.624 (dashed horizontal line). In early iterations, $d_{12}^{(t)}$ frequently takes large values, indicating strong exploration of the direction space, while later iterations stabilize at a lower yet consistently elevated distance as the load path selection focuses on uncertain regions without collapsing to redundant queries. While **Figure 2(a)** focuses on intra-batch directional diversity, **Appendix A** reports a complementary global redundancy check, confirming that the actively acquired directions do not become globally redundant. **Figure 2(b)** shows the evolution of the mean committee disagreement over the active-learning iterations, quantified by the variance of committee predictions at the selected queries (mean over three runs; shaded band: ±1 standard deviation). A pronounced reduction in variance is observed during the early iterations, reflecting the rapid elimination of high-uncertainty regions in stress space as informative loading directions are incorporated into the training set. After approximately 30–40 iterations, the variance decreases more gradually and stabilizes near zero, indicating increasing agreement among committee members. Importantly, this monotonic reduction in stress-space variance occurs alongside the elevated directional diversity shown in **Figure 2(a)**, demonstrating that uncertainty reduction is achieved without sacrificing directional exploration. The batch-mode strategy, therefore, simultaneously enforces directional diversity on $S^5$ and reduces predictive uncertainty in the stress space.

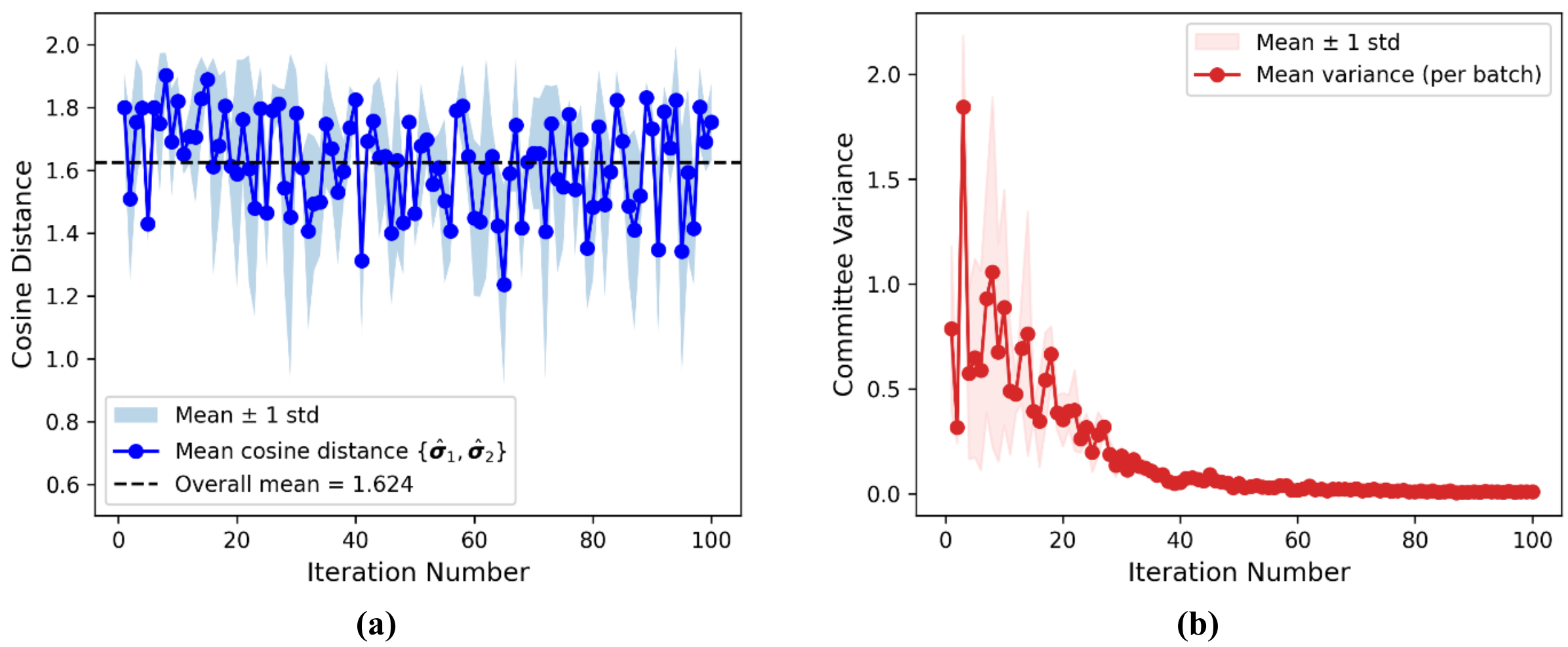

**(a)** **(b)**

**Figure 2 –** Intra-batch diversity and uncertainty reduction for batch size $b = 2$, averaged over three independent runs. The shaded regions indicate ±1 standard deviation **(a).** Intra-batch diversity is shown as the cosine distance between the first two selected directions $\hat{\boldsymbol{\sigma}}_1$, $\hat{\boldsymbol{\sigma}}_2$ (blue). The dashed line denotes the overall mean distance (1.626), well above the random-pair baseline. **(b)** shows the committee disagreement at the selected queries, quantified by the prediction variance, which is plotted versus iteration.

**Figure 3** compares batch-mode active learning with batch size $b = 2$ against a sequential baseline (variance-only), using the MCC on a held-out test set. Curves show the mean over three independent runs, with shaded bands indicating ±1 standard deviation. The two panels separate two different notions of efficiency: oracle-query efficiency (performance per oracle call) and update efficiency (performance per retraining cycle, serving here as a proxy metric for time-to-solution). **Figure 3(a)** plots MCC as a function of the total number of oracle-queried directions. The batch-2 curve ( purple) closely tracks the sequential baseline (orange), indicating that batching does not degrade sample efficiency. For a given number of queried samples, both strategies reach similar MCC values and converge to essentially the same performance plateau. This indicates that selecting two points per retraining cycle does not compromise the quality of the acquired training data, and batching does not waste oracle calls.

**Figure 3(b)** presents the same results as a function of iteration budget (number of retraining cycles). In the present setup, retraining dominates computational cost, as it involves training SVC models together with independent hyperparameter optimization for each committee member. Because this cost is incurred once per iteration, irrespective of batch size, the iteration count provides a proxy metric for wall-clock time. Direct runtime measurements support this interpretation: under identical hardware and software conditions, the sequential strategy required approximately 6 hours to complete the active-learning process, whereas batch-mode sampling with batch size 2 required approximately 3 hours on the same machine (4 cores, 8 logical processors and 8 GB RAM), closely matching the factor-of-two reduction in retraining cycles. Under this cost model, batch-2 achieves higher MCC for a fixed iteration budget, particularly in the early and intermediate stages of learning. This reflects its ability to acquire two informative samples per retraining cycle, effectively doubling the amount of information obtained per expensive model update without sacrificing learning quality. Taken together, the two panels demonstrate that batch size 2 preserves almost the same sample efficiency of sequential active learning while substantially reducing the number of costly retraining cycles required to reach a given level of predictive performance.

These results highlight the main practical advantage of batch-mode active learning: it preserves oracle-query efficiency while improving time-to-solution by reducing the number of costly retraining updates. In our implementation, retraining dominates the runtime, so batching primarily reduces wall-clock time by cutting the number of retraining cycles. In simulation-driven settings where evaluations are computationally expensive and can be executed in parallel (e.g., multiple independent simulations on HPC resources), batching offers an additional (and potentially larger) reduction in time by enabling concurrent oracle queries within each iteration.

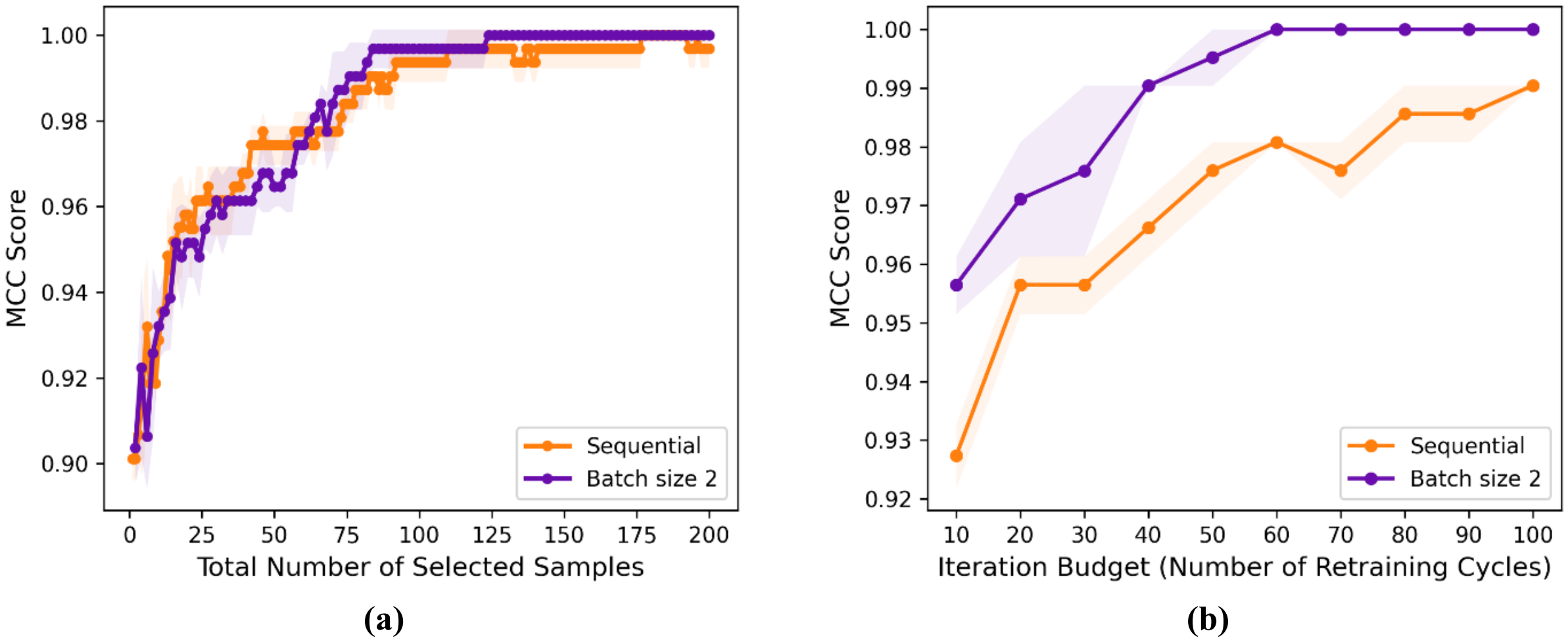

**Figure 3** – Comparison of predictive performance of sequential active learning versus batch-mode active learning ($b = 2$) measured by MCC on a held-out test set (mean over three runs; shaded band: ±1 standard deviation). **(a)** MCC versus total number of queries. For a fixed query budget, batch-mode sampling (purple) closely tracks the sequential baseline (orange) and reaches nearly the same performance plateau. **(b)** MCC versus iteration budget (number of retraining cycles). Because each iteration corresponds to a full retraining step and retraining cost is incurred once per iteration, batch-mode sampling attains higher MCC for a given iteration budget, demonstrating improved efficiency per (expensive) model update.

To illustrate the geometric evolution of the learned yield surface, **Figure 4** visualizes the ML yield function on the π-plane (deviatoric plane in principal stress space) using cylindrical coordinates $(\theta, \sigma_{\text{eq}})$. In addition to the binary class prediction ($-1$ for elastic and $+1$ for plastic), the classifier returns a signed decision value, which is used to generate the background color map. Negative values (purple) correspond to elastic classification, whereas positive values (brown) correspond to plastic classification. While the absolute magnitude of this decision value has no direct physical meaning, it provides a signed measure of the location relative to the classification boundary, and its zero-level set defines the learned yield locus. **Figure 4(a)** shows the yield locus obtained from the initial training set consisting of 100 randomly sampled unit loading directions. Although the general anisotropic shape is already captured, noticeable deviations from the reference Hill yield surface remain in certain angular regions, reflecting incomplete coverage of stress space by purely random sampling. **Figure 4(b)** presents the reconstructed yield locus after 100 active learning iterations. The learned zero-level set closely matches the reference yield surface over the full angular range, demonstrating that adaptive sampling progressively improves stress-space coverage and refines the yield-surface boundary.

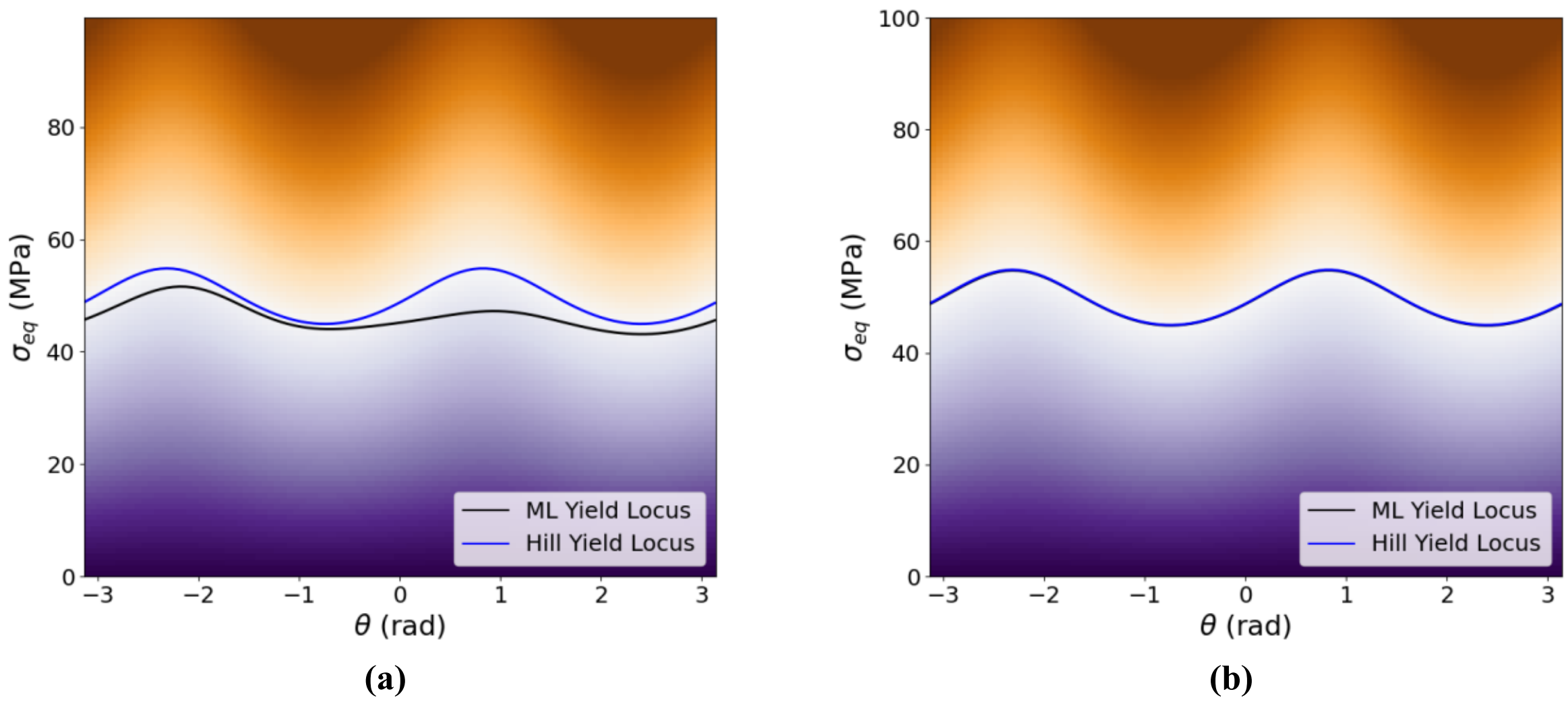

**Figure 4** – Evolution of the learned yield surface on the π-plane. The yield function is visualized in cylindrical coordinates $(\theta, \sigma_{\text{eq}})$. where $\theta$ denotes the polar angle on the deviatoric (π) plane and $\sigma_{\text{eq}}$ is the equivalent stress. The color map represents the signed decision values of the SVC: negative values (purple) correspond to elastic classification, positive values (brown) to plastic classification, and the yield locus is given by the zero contour $f(\boldsymbol{\sigma}) = 0$. **(a)** Initial SCV model trained on 100 randomly sampled unit loading directions before applying active learning. **(b)** SVC model after 100 active-learning iterations (batch size 2).

## 3.2. Evaluation of Larger Batch Sizes

To assess whether the proposed batch-mode sampling strategy remains effective for larger batch sizes, we extend the analysis beyond batch size 2 and consider batch sizes of 3 and 4. As for batch 2, all reported metrics are averaged over three independent runs, and shaded bands indicate ±1 standard deviation. **Figure 5** summarizes within-batch diversity and the associated reduction in committee disagreement for these larger batches. **Figure 5(a)** extends the intra-batch diversity analysis to batch size 3. As in the batch-size-2 case, the first direction $\hat{\boldsymbol{\sigma}}_1{}^{(t)}$ is selected by maximizing committee disagreement (Eq. (7)), and the remaining directions are obtained using the diversity-aware objective (Eq. (10)). To quantify diversity as the batch is processed, in each iteration $t$, we report (i) the cosine distance between the first two selected directions $\hat{\boldsymbol{\sigma}}_1{}^{(t)}$ and $\hat{\boldsymbol{\sigma}}_2{}^{(t)}$ as $d_{12}^{(t)} = 1 - \hat{\boldsymbol{\sigma}}_1{}^{(t)} \cdot \hat{\boldsymbol{\sigma}}_2{}^{(t)}$ shown by the blue curve, and (ii) the mean cosine distance of the third direction $\hat{\boldsymbol{\sigma}}_3{}^{(t)}$ from the first two, $d_{3|\{1,2\}}^{(t)} = \frac{1}{2}\left[\left(1 - \hat{\boldsymbol{\sigma}}_3{}^{(t)} \cdot \hat{\boldsymbol{\sigma}}_1{}^{(t)}\right) + \left(1 - \hat{\boldsymbol{\sigma}}_3{}^{(t)} \cdot \hat{\boldsymbol{\sigma}}_2{}^{(t)}\right)\right]$ shown by the orange curve. The $d_{12}^{(t)}$ values remain consistently high, indicating that the diversity term successfully prevents redundant selections early within each batch. Once $\hat{\boldsymbol{\sigma}}_1{}^{(t)}$ is fixed (chosen by maximizing the variance), there is sufficient geometric freedom on $S^5$ to select $\hat{\boldsymbol{\sigma}}_2{}^{(t)}$ with strong angular diversification relative to it. In contrast, $d_{3|\{1,2\}}^{(t)}$ values are smaller because the third direction has to be diversified with respect to the two already chosen directions at the same time which imposes a stronger geometric constraint on $S^5$. **Figure 5(b)** shows the corresponding evolution of the mean committee disagreement. Despite the stronger within-batch constraint introduced by selecting three directions per iteration, the variance drops rapidly in the early iterations and then levels off to a low plateau. This indicates that the batch-3 acquisition continues to reduce model uncertainty while maintaining meaningful intra-batch diversity.

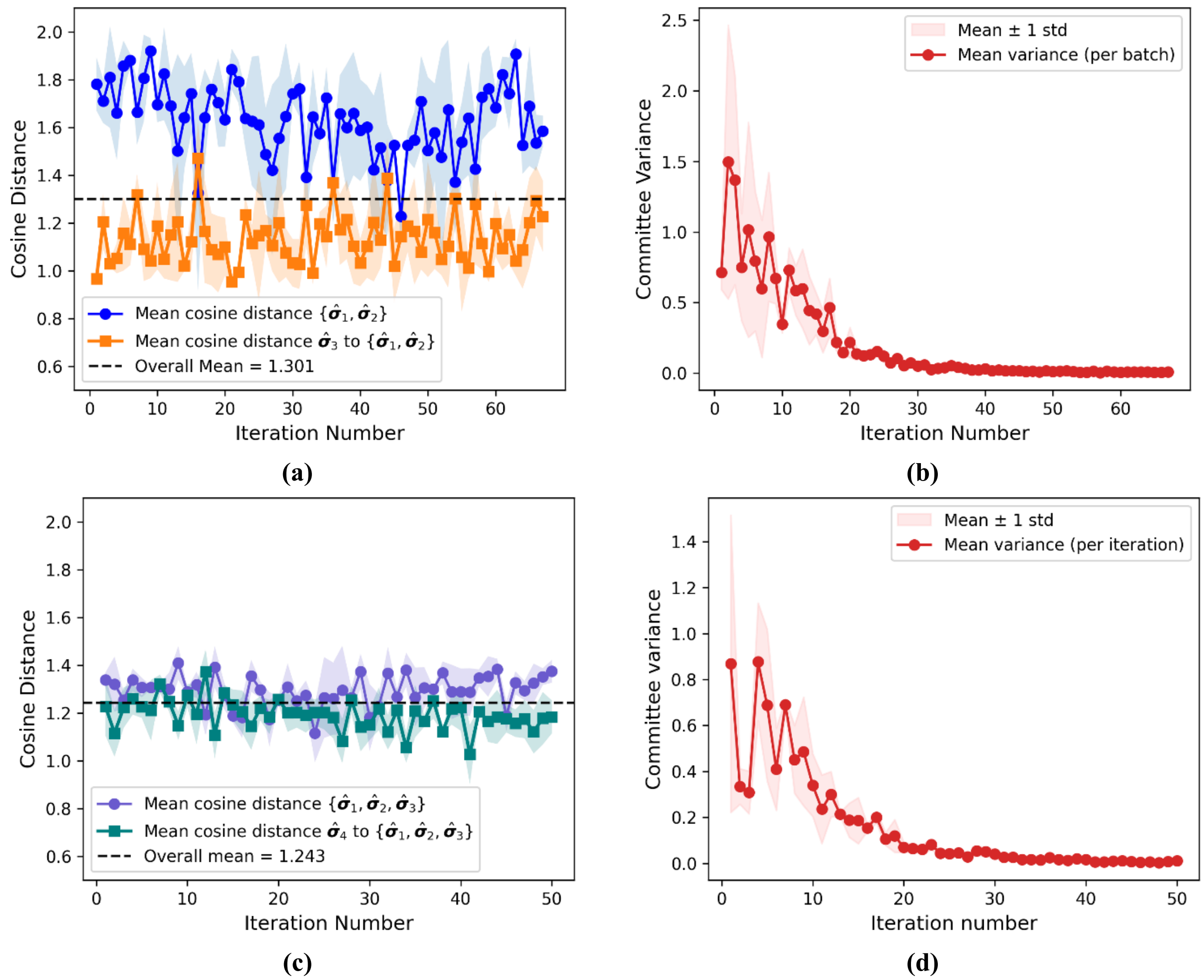

**Figure 5** – Intra-batch diversity and uncertainty reduction for batch sizes $b = 3$ and $b = 4$ averaged over three independent runs. The shaded regions indicate ±1 standard deviation. **(a)** $b = 3$: intra-batch diversity shown as the cosine distance between the first two selected directions $\hat{\boldsymbol{\sigma}}_1$, $\hat{\boldsymbol{\sigma}}_2$ (blue) and the mean cosine distance of the third direction $\hat{\boldsymbol{\sigma}}_3$ to $\{\hat{\boldsymbol{\sigma}}_1, \hat{\boldsymbol{\sigma}}_2\}$ (orange). **(b)** $b = 3$: evolution of the mean committee disagreement over iterations. **(c)** $b = 4$: intra-batch diversity measured by the mean pairwise cosine distance among $\{\hat{\boldsymbol{\sigma}}_1, \hat{\boldsymbol{\sigma}}_2, \hat{\boldsymbol{\sigma}}_3\}$ (purple) and the mean cosine distance of $\hat{\boldsymbol{\sigma}}_4$ to $\{\hat{\boldsymbol{\sigma}}_1, \hat{\boldsymbol{\sigma}}_2, \hat{\boldsymbol{\sigma}}_3\}$ (dark green). **(d)** $b = 4$: evolution of the mean committee disagreement over iterations. The dashed black line in **(a)** and **(c)** denotes the overall mean cosine distance.

**Figure 5(c)** extends the intra-batch diversity analysis to batch size 4. As before, the first direction $\hat{\boldsymbol{\sigma}}_1{}^{(t)}$ is selected by maximizing committee variance (Eq. (7)), and the remaining directions are added sequentially using the diversity-aware criterion (Eq. (10)). For clarity, we report two summary metrics that capture the key behavior as the batch is processed. First, we compute the mean pairwise cosine distance among the first three selected directions, $d^{(t)}_{\{1,2,3\}} = \frac{1}{3}\left[\left(1 - \hat{\boldsymbol{\sigma}}_1{}^{(t)} \cdot \hat{\boldsymbol{\sigma}}_2{}^{(t)}\right) + \left(1 - \hat{\boldsymbol{\sigma}}_1{}^{(t)} \cdot \hat{\boldsymbol{\sigma}}_3{}^{(t)}\right) + \left(1 - \hat{\boldsymbol{\sigma}}_2{}^{(t)} \cdot \hat{\boldsymbol{\sigma}}_3{}^{(t)}\right)\right]$, shown by the purple curve. Second, we quantify how the fourth direction diversifies relative to the already selected triplet by its mean cosine distance to $\left\{\hat{\boldsymbol{\sigma}}_1{}^{(t)}, \hat{\boldsymbol{\sigma}}_2{}^{(t)}, \hat{\boldsymbol{\sigma}}_3{}^{(t)}\right\}$, $d^{(t)}_{4|\{1,2,3\}} = \frac{1}{3}\left[\left(1 - \hat{\boldsymbol{\sigma}}_4{}^{(t)} \cdot \hat{\boldsymbol{\sigma}}_1{}^{(t)}\right) + \left(1 - \hat{\boldsymbol{\sigma}}_4{}^{(t)} \cdot \hat{\boldsymbol{\sigma}}_2{}^{(t)}\right) + \left(1 - \hat{\boldsymbol{\sigma}}_4{}^{(t)} \cdot \hat{\boldsymbol{\sigma}}_3{}^{(t)}\right)\right]$, shown by the dark green curve. The first metric reflects the mutual diversity achieved during the early part of batch construction, while the second measures the marginal diversification contributed by the final pick. The $d^{(t)}_{\{1,2,3\}}$ values remain stable and relatively high across iterations, indicating that the diversity term continues to disfavor redundant directions as the batch grows.

In contrast, $d^{(t)}_{4|\{1,2,3\}}$ is smaller, since the fourth pick must be simultaneously diversified with respect to three existing directions, which is inherently more restrictive than diversifying against one or two directions. Nevertheless, $d^{(t)}_{4|\{1,2,3\}}$ remains close to the overall mean level, showing that the final selection still contributes to meaningful additional diversification. **Figure 5(d)** shows the corresponding evolution of the mean committee disagreement for batch size 4. Despite the stronger within-batch diversity constraint imposed by selecting four directions per iteration, the variance decreases rapidly in the early iterations and then approaches a low plateau. This indicates that the batch-4 strategy remains effective at acquiring informative samples that reduce prediction uncertainty while maintaining directional diversity within each batch. It should be noted that **Figures 5(a)–(c)** quantify intra-batch directional diversity for batch sizes 3 and 4. Complementary global redundancy checks, are reported in **Appendix A**. This analysis proves that the actively acquired directions, even in higher batch sizes, do not become globally redundant.

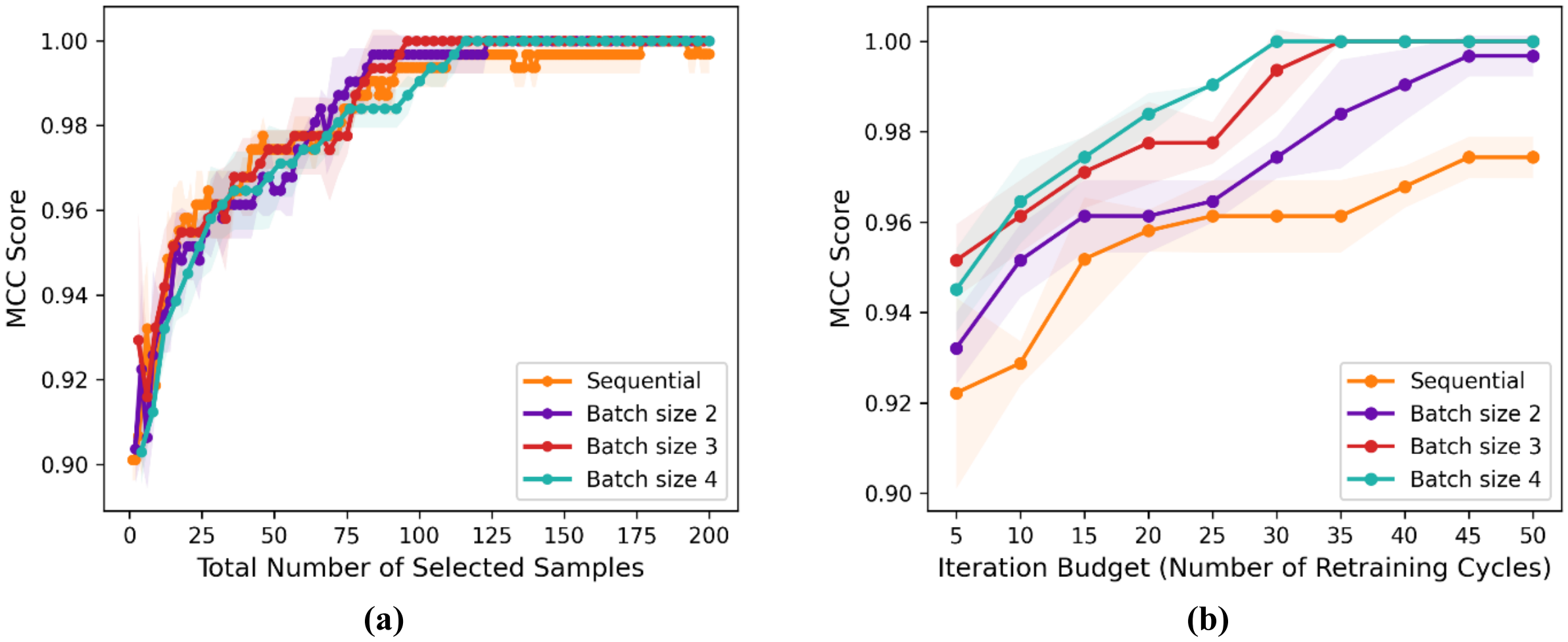

**Figure 6** – Effect of batch size on active-learning efficiency evaluated for MCC on a held-out test set for the sequential variance-only baseline and batch-mode sampling $b = 2, 3$, and 4 (mean over three independent runs with shaded bands showing ±1 standard deviation.) **(a)** MCC versus total number of queries. **(b)** MCC versus iteration budget (number of retraining cycles), where each iteration corresponds to one full committee retraining step.

**Figure 6** extends the efficiency comparison to batch-mode sampling with batch sizes $b = 2, 3,$ and $4,$ against the fully sequential baseline, using MCC on a held-out test set for evaluation. The resulting curves show the mean over three independent runs, with shaded bands indicating ±1 standard deviation. As in **Figure 3**, the two panels separate query efficiency (performance per oracle call) and update efficiency (performance per retraining cycle, used here as a proxy metric for time-to-solution). **Figure 6(a)** plots MCC versus the total number of queried load directions. Across batch sizes, batch-mode acquisition reaches almost the same performance plateau as the sequential baseline, indicating that increasing the batch size has no strong effect on query efficiency in our study. For a fixed labeling budget, batch sizes 3 and 4 achieve MCC values comparable to sequential acquisition, with only modest differences during the early stages that fall within run-to-run variability. **Figure 6(b)** presents the same results as a function of iteration budget (number of retraining cycles), where each iteration corresponds to a full committee retraining step. Since retraining (including hyperparameter optimization for each committee member) dominates computational cost, in our study, and is incurred once per iteration regardless of batch size, the iteration count provides a practical proxy metric for wall-clock time. Under this cost model, larger batches yield clear update-efficiency gains: $b = 3$ and $b = 4$ achieve substantially higher MCC than the sequential baseline for the same number of retraining cycles, especially in the early and intermediate stages. Consistent with this trend, $b = 4$ reaches near-peak performance in the fewest retraining cycles, followed by $b = 3$ and then $b = 2$. Taken together, the two figures show that larger batch sizes preserve query efficiency while reducing the number of costly retraining cycles required to achieve a given predictive performance level.

Summarized, in the present study, Hill's anisotropic yield criterion is used as a benchmark oracle to generate ground-truth labels efficiently and to analyze and assess the behavior of the proposed sampling strategy. In our current implementation, oracle queries are inexpensive, and the dominant cost is model retraining (committee training and hyperparameter optimization), so the observed speedup primarily reflects the reduction in retraining cycles. In practical simulation-driven settings, however, each queried stress state may require a full high-fidelity simulation (and associated postprocessing and data handling), so oracle evaluations can become a major (sometimes dominant) contributor to wall-clock time. In that regime, batch-mode acquisition can be beneficial on two fronts: it reduces the number of costly model updates, and it enables multiple oracle evaluations to be executed concurrently within each iteration. Compared with a fully sequential workflow, where each iteration runs a single simulation and then retrains before selecting the next query, batching reduces the number of sequential round trips and, when parallel resources are available, can substantially increase time-efficiency.

Batch sizes 3 and 4 largely preserve oracle-query efficiency relative to sequential acquisition while reducing the number of expensive retraining cycles required to reach a given MCC. We do not pursue larger batch sizes because the batch-construction problem becomes increasingly constrained: as more points must be selected simultaneously, it becomes harder to maintain both high uncertainty and strong within-batch diversity, increasing the risk of redundancy. This represents a certain limitation of our strategy. Moreover, larger batches reduce the frequency of model updates, which can delay correction of model bias and slow refinement in later stages of learning. Based on this trade-off, we consider $b = 4$ a practical upper bound for the present study.

## 4. Conclusions

Efficient training dataset generation is a key challenge in data-driven constitutive modeling, particularly when constitutive responses must be learned from six-dimensional stress spaces. In this work, we propose a diversity-aware batch-mode query-by-committee active learning strategy for adaptive sampling, motivated by the need to reduce redundant evaluations while maintaining informative coverage of the feature space. Although the approach is general in scope, it was benchmarked here for training machine learning yield functions in six-dimensional stress space, where the elastic-plastic boundary is approximated a support vector classifier. In this setting, efficient sampling is essential because training quality depends strongly on how the stress space is explored, while repeated retraining and ground-truth evaluations can become costly. The proposed method addresses this by adopting a query-by-committee strategy, in which a committee of such classifiers is used to identify the most informative loading directions through predictive variance, while an additional diversity criterion avoids redundant selections within each batch. This combination enables efficient exploration of stress space while reducing retraining effort relative to sequential active learning approaches.

In this benchmark study, the queried samples are unit loading directions in the six-dimensional stress space, while Hill's anisotropic yield criterion is used as the reference material model to determine the corresponding yield onset and generate ground-truth data with a negligible numerical effort. Within this setting, the proposed active-learning strategy is evaluated for different batch sizes. The cosine-similarity-based analysis of sampled loading directions shows that a high within-batch diversity is maintained throughout the active-learning iterations. At the same time, the method systematically reduces committee uncertainty while continuing to explore distinct regions of stress space. The results further show that batch-mode sampling preserves query efficiency relative to fully sequential data acquisition, while providing clear gains in update efficiency when retraining is costly. Consequently, larger batches achieve comparable or higher prediction accuracy in fewer retraining cycles.

The numerically highly efficient ground-truth data generation allows us to assess the effect of the sampling strategy on isolation. In this benchmark setting, batch-mode acquisition improves efficiency primarily by reducing the number of costly retraining cycles, which dominate the wall-clock time. In practical applications, however, such ground-truth data would typically be obtained from computationally expensive high-fidelity simulations. In that case, ground-truth data generation may itself become the dominant cost if the required simulations are performed sequentially. However, when simulations and postprocessing can be parallelized, batch-mode sampling can provide substantially larger wall-clock savings than observed in the present benchmark.

**Acknowledgements**

Lukas Morand and Dirk Helm acknowledge support from the European Commission under the European Union's Horizon Research and Innovation program (Grant Agreement No. 101091912). Views and opinions expressed are, however, those of the author(s) only and do not necessarily reflect those of the European Union. Neither the European Union nor the granting authority can be held responsible for them.

**Data availability statement**

The benchmark dataset supporting the findings of this study is publicly available on Zenodo (DOI: 10.5281/zenodo.19666411).

**Author contributions**

R. Shoghi and L. Morand contributed equally to conceptualization and methodology. R. Shoghi led data generation and analysis and wrote the original draft. L. Morand supported the analysis and contributed to manuscript review and editing. A. Hartmaier supported conceptualization, methodology, and analysis, contributed to review and editing, and led supervision. D. Helm contributed to the review and editing and supported supervision.

**Declaration of generative AI and AI-assisted technologies in the writing process**

During the preparation of this work, the author(s) used LLM tools to improve the readability and English language of the manuscript. After using this tool/service, the authors reviewed and edited the content as needed and took full responsibility for the content of the published article.

# Appendix A: Redundancy analysis for batch-mode sampling

The presented strategy for batch-mode active learning can become inefficient, if load paths selected within a batch (or across batches) become too similar, i.e., if the algorithm repeatedly picks nearly identical loading directions. Because our method explicitly encourages within-batch diversity, we include here a simple redundancy check to confirm that the actively acquired directions remain largely non-duplicative. Starting from batch size $b = 2$, for each of three independent runs, we collect the actively selected unit directions (200 samples, corresponding to indices 101–300) and compute the full cosine similarity matrix $M$, where $M_{jk} = \hat{\boldsymbol{\sigma}}_j^{\mathrm{T}} \hat{\boldsymbol{\sigma}}_k \in [-1,1]$. Values close to 1 indicate nearly parallel directions (potential redundancy), while values close to 0 indicate near-orthogonality, and negative values indicate opposing directions. **Figure A1** visualizes $M$ for the three runs. The diagonal is identically 1 (self-similarity). Redundancy would manifest as large off-diagonal red blocks or extended red bands, indicating many pairs of directions with cosine similarity close to 1. Instead, the off-diagonal entries are dominated by near-zero and negative values, with only sparse isolated high-similarity pairs. This qualitative pattern indicates that the selected directions remain broadly spread in stress space and do not collapse into repeated or near-duplicate queries.

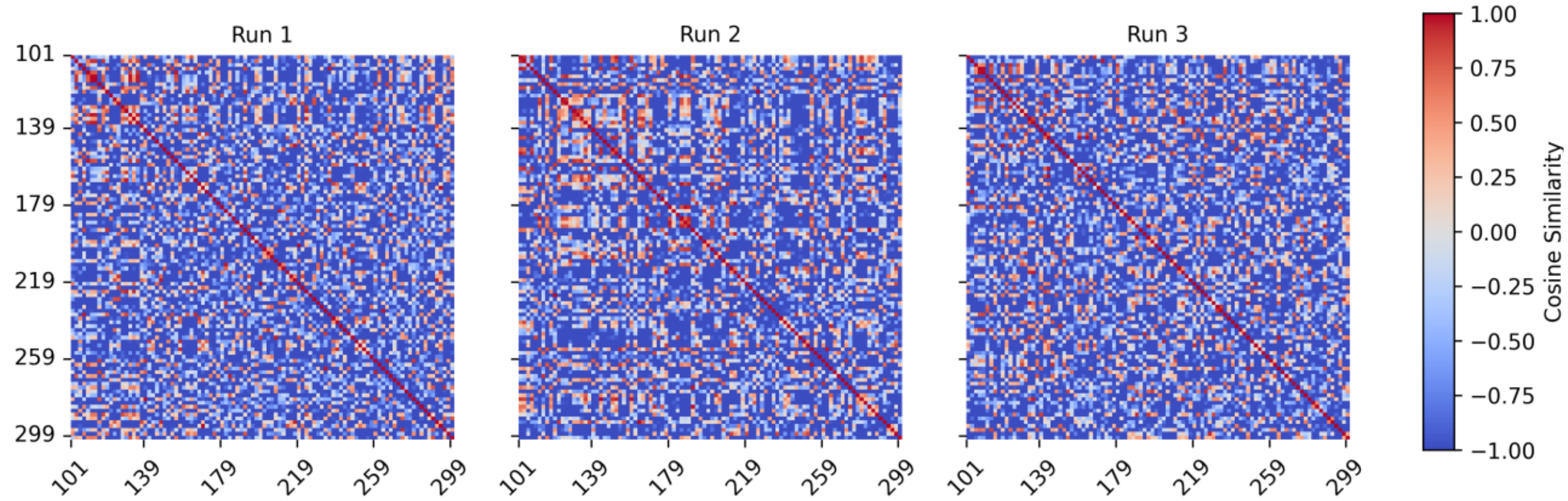

**Figure A1 –** Pairwise cosine similarity matrices of the actively selected unit loading directions for three independent runs of batch-mode active learning with batch size $b = 2$ (200 active samples; indices 101–300). Colors indicate cosine similarity $[-1,1]$

**Table A1** provides quantitative support for this observation. The mean off-diagonal similarity is almost zero across runs $(-0.0043, -0.0031, -0.0040)$, indicating no systematic tendency toward positively aligned (similar) directions. The mean pairwise cosine distance is high $(0.969 - 0.984)$, consistent with generally well-separated directions. Most importantly, the fraction of direction pairs with a cosine similarity of 0.90 or higher remains small, ranging from $1.88\%$ and $2.67\%$, showing that near-duplicate pairs are rare. Overall, Fig. A1 and Table A1 confirm that, for batch size $b = 2$, the batch-mode data acquisition strategy produces a largely non-redundant set of loading directions, with only a small number of highly similar pairs.

**Table A1 –** Redundancy statistics for batch-mode active learning with batch size $b = 2$, computed over the actively selected unit loading directions (200 active samples; indices 101–300) for three independent runs. Reported are the mean and standard deviation of off-diagonal cosine similarities, the mean pairwise cosine distance, and the fraction of pairs with cosine similarity ≥ 0.90.

| Run | Mean off-diagonal similarity | Std-off- diagonal similarity | Mean pairwise distance | % pairs ≥ 0.90 similarity |
|---|---|---|---|---|
| 1 | -0.004300 | 0.498293 | 0.968925 | 0.018788 |
| 2 | -0.003116 | 0.521092 | 0.984312 | 0.026667 |

| 3 | -0.004022 | 0.498758 | 0.976526 | 0.019596 |
|---|---|---|---|---|

Batch-mode redundancy becomes more critical as the batch size increases, because more points are selected before the model is updated. With a larger batch, each additional point must be chosen to be informative and sufficiently different from multiple already selected directions in the same iteration. This makes the data acquisition rule prone to crowd into similar regions and produce near-duplicates. To assess whether this effect becomes problematic at larger batch sizes, we repeat the same redundancy analysis for batch size $b = 4$ and report the corresponding summary statistics in Table A2 (three independent runs).

**Table A2** – Redundancy statistics for batch-mode active learning with batch size $b = 4$, computed over the actively selected unit loading directions (200 active samples; indices 101–300) for three independent runs. Reported are the mean and standard deviation of off-diagonal cosine similarities, the mean pairwise cosine distance, and the fraction of pairs with cosine similarity $\geq 0.90$.

| Run | Mean off-diagonal similarity | Std-off- diagonal similarity | Mean pairwise distance | % pairs ≥ 0.90 similarity |
|---|---|---|---|---|
| 1 | -0.003577 | 0.491792 | 1.003577 | 0.016231 |
| 2 | -0.002866 | 0.504674 | 1.002866 | 0.019447 |
| 3 | -0.004427 | 0.510581 | 1.004427 | 0.016935 |

Compared to batch size $b = 2$ (Table A1), the results indicate that redundancy remains limited. The mean off-diagonal cosine similarity stays close to zero indicating no systematic tendency toward selecting similar directions. The standard deviation of off-diagonal similarities is also comparable ($\approx 0.49$ to $0.51$), suggesting a similar overall spread of pairwise angles. I Importantly, the fraction of direction pairs with cosine similarity of 0.90 or higher remains low for batch size $b = 4$ $(1.62\%$ to $1.94\%)$ and is not elevated relative to batch size $b = 2$ $(1.88\%$ to $2.67\%)$. Finally, the mean pairwise cosine distance is approximately $1.003$ for batch size $b = 4$, indicating that the selected directions remain close to orthogonal on average. Taking together, these results show that even in higher batch sizes, the diversity mechanism continues to prevent widespread near-duplicate querying, and using batch-mode does not lead to a noticeable increase in redundancy.